\documentclass[twocolumn]{aa}

\usepackage{epsfig}
\usepackage{amsmath}
\usepackage{amssymb}
\usepackage{natbib}
\usepackage{subfigure}
\usepackage{longtable}
\usepackage{graphicx}
\usepackage{textcomp}
\usepackage{color}
\usepackage{url}
\begin{document} %
\def\simlt{\mathrel{\rlap{\lower 3pt\hbox{$\sim$}}\raise 2.0pt\hbox{$<$}}}
\def\simgt{\mathrel{\rlap{\lower 3pt\hbox{$\sim$}} \raise
2.0pt\hbox{$>$}}}
\def\lsim{\,\lower2truept\hbox{${<\atop\hbox{\raise4truept\hbox{$\sim$}}}$}\,}
\def\gsim{\,\lower2truept\hbox{${> \atop\hbox{\raise4truept\hbox{$\sim$}}}$}\,}

\title{Black-hole mass estimates for a homogeneous sample of bright flat-spectrum radio quasars}
   \titlerunning{Black hole masses of flat spectrum radio quasars}
\authorrunning{G. Castignani et al.}
   \author{G. Castignani
          \inst{1}\fnmsep\thanks{e-mail: castigna@sissa.it}
                 \and
        F. Haardt\inst{2,3}
                  \and
          A. Lapi\inst{4,1}
  \and
          G. De Zotti\inst{5,1}
           \and
          A. Celotti\inst{1}
            \and
          L. Danese\inst{1}
          }

   \institute{SISSA, Via Bonomea 265, 34136, Trieste, Italy
         \and
 DiSAT, Universit\`a dell'Insubria, via Valleggio 11, I-22100 Como, Italy
 \and
 INFN, Sezione di Milano Bicocca, Piazza Della Scienza 3, I-20126 Milano
         \and
         Dipartimento di Fisica, Universit\`a `Tor Vergata', Via della Ricerca Scientifica 1, I-00133 Roma, Italy
         \and
          INAF-Osservatorio Astronomico di Padova, Vicolo dell'Osservatorio 5, I-35122 Padova, Italy
             }

   \date{March 6th, 2013}


\abstract
{We have selected a complete sample of flat-spectrum radio quasars (FSRQs) from the WMAP 7-yr
catalog within the SDSS area, all with measured redshift, and have compared the black hole
mass estimates based on fitting a standard accretion disk model to the `blue bump' with those
 obtained from the commonly used single epoch virial method. The sample comprises 79 objects
with a flux density limit of 1 Jy at 23 GHz, 54 of which (68\%) have a clearly detected `blue bump'.
 Thirty-four of the latter have, in the literature, black hole mass estimates obtained with the virial method.
 The mass estimates obtained from the two methods are well correlated. If the calibration factor of the virial
 relation is set to $f=4.5$, well within the range of recent estimates, the mean logarithmic ratio of the two mass estimates is equal to zero with a dispersion close to the estimated uncertainty of the virial method. The fact that the two independent methods agree so closely in spite of the potentially 
large uncertainties associated with each lends strong support to both of them. The distribution of black-hole masses 
for the 54 FSRQs in our sample with a well detected blue bump has a median value of $7.4\times 10^{8}\,M_\odot$. 
It declines at the low mass end, consistent with other indications that radio loud AGNs are generally associated with the most massive black holes, although the decline may be, at least partly, due to the source selection. The distribution drops above $\log(M_\bullet/M_\odot) = 9.4$, implying 
that ultra-massive black holes associated with FSRQs must be rare.}

\keywords{galaxies: active -- quasar: general -- black hole physics}

\maketitle

\section{Introduction}\label{par:intro}

Reliable mass estimates of black-holes (BHs) in active galaxies are essential to investigate the physics of accretion and emission processes in the BH environment and the link between the BH growth and the evolution of galaxy stellar populations. 
However, getting them is not easy however.
Dynamical mass estimates are only
possible for nearby objects whose parsec-scale BH sphere of influence can be
resolved, and are usually applicable to quiescent galaxies.
BH masses of luminous active galactic nuclei (AGNs)
are most commonly estimated using a technique known as `single epoch virial method' or, briefly, `SE method'. Under the usual assumption that the broad-line region (BLR) is in virial equilibrium the BH mass is derived as
\begin{equation}
 \label{eq:virial_BHmass}
 M_{\bullet} = f\frac{R_{\rm BLR}\Delta V^2}{G},
\end{equation}
where $R_{\rm BLR}$ is the BLR radius, $\Delta V$ is the velocity of the BLR clouds (that can be inferred from the line width), $f$ is the virial coefficient that depends on the geometry and kinematics of the BLR, and $G$ is the gravitational constant. An effective way to estimate $R_{\rm BLR}$, known as `reverberation mapping', exploits the delay in the response of the BLR to short-term variability of the ionizing continuum \citep{BlandfordMcKee1982}. The application of this technique has however been limited because it requires long-term monitoring of both the continuum and the broad emission lines. The SE method bypasses this problem exploiting the correlation between the size of the BLR and the AGN optical/UV continuum luminosity empirically found from reverberation mapping \citep{KoratkarGaskell1991,Kaspi2005,Bentz2009a} and expected from the photo-ionization model predictions \citep{KoratkarGaskell1991}. The AGN continuum luminosity can thus be used as a proxy for the BLR size.

However, measurements of the AGN continuum may be affected by various systematics: contributions from broad Fe\,{\sc II} emission and/or from the host galaxy, and, in the case of blazars, contamination by synchrotron emission from 
the jet \citep{Wu2004,GreeneHo2005}. Fortunately, there are tight, almost linear correlations between the luminosity of the AGN continuum and the luminosity of emission lines such as H$\alpha$, H$\beta$, Mg\,{\sc II} and C\,{\sc IV} \citep{GreeneHo2005,VestergaardPeterson2006,Shen2011}. It is thus expedient to estimate the BH masses using the line luminosities and full widths at half maximum (FWHMs).

Yet the reliability and accuracy of the method and of the resulting mass estimates, $M_{\bullet}$, is debated  \citep{Croom2011,Assef2012}. Each of its ingredients is endowed with a considerable uncertainty \citep{VestergaardPeterson2006,Park2012b}. Recent estimates of the virial coefficient, $f$, differ by a factor $\simeq 2$. The luminosity-size relations have a significant scatter. In addition line-widths and luminosities vary on short timescales while black hole masses should not vary. These uncertainties are on top of those on the measurements of line-widths and luminosities, that need to be corrected for emissions from outside the BLR. \citet{Park2012b} found that uncertainties in the size--luminosity relation and in the virial coefficient translate in a factor $\simeq 3$ uncertainty in $M_\bullet$.
But, as pointed out by \citet{Shen2013}, other sources of substantial systematic errors may also be present.

An independent method to estimate $M_\bullet$ rests upon fitting the optical/UV `bump' of AGNs
\citep[e.g.][]{Malkan1983,Wandel_Petrosian1988,Laor1990,Ghisellini2010,Calderone2012}.
In the the \citet{ShakuraSunyaev1973} accretion disk model
the BH mass is a simple function of the frequency at which the disk emission peaks,
which is a measure of the effective disk temperature, of the accretion rate,
estimated by the disk luminosity, given the radiation efficiency and the inclination angle,
$i$, i.e. the angle between the line-of-sight and the normal
to the disk plane \citep{Frank2002}.
However, this method had a limited application to estimate $M_\bullet$ \citep{FerrareseFord2005}
mainly because reliable estimates of the intrinsic disk luminosity are very difficult to obtain.
In fact: a) the inclination is generally unknown and the observed flux density is proportional to $\cos{i}$;
 b) the observed UV bump is highly sensitive to obscuration by dust either in the circum-nuclear torus or in the host galaxy;
c) we need to subtract the contribution from the host galaxy that may be substantial particularly for the weaker active nuclei.

These difficulties are greatly eased in the case of flat-spectrum radio quasars (FSRQs) because: a) the accretion disk is expected to be perpendicular to the jet direction, and indeed there is good evidence that the jets of {\it Fermi} FSRQs are highly aligned (within $5^\circ$) with the line-of-sight \citep{Ajello2012} so that $\cos{i}\simeq 1$; b) the obscuration is expected to be negligible because blazar host galaxies are thought to be passive, dust free, ellipticals \citep[e.g.][and references therein]{Giommi2012} and also the torus is likely perpendicular to the line-of-sight; c) the contamination is also small because elliptical hosts are red, i.e. are faint in the UV. However, the UV emission may be contaminated by the emission from the relativistic jet.

On the other hand, the BH mass estimates obtained by fitting the blue bump rely on several assumptions whose validity is not fully proven \citep[see][for a discussion]{Ghisellini2010}: (i) the disk is described by a standard \citet{ShakuraSunyaev1973} model, i.e. the disk is optically thick and geometrically thin; (ii) the black hole is non-rotating, of Schwarzschild type; (iii) the SED is a combination of black body spectra. If any of these assumptions does not hold, the mass estimates would be affected. An additional uncertainty source  in our practical application is that photometry of the SED beyond the peak is not always available. When it is available, it is not simultaneous with the optical data determining the rising part of the SED and needs corrections for extinction within our own galaxy and, in the case of objects at high-$z$, for photoelectric absorption in the intergalactic medium.

A cross-check of the outcome of the two, independent, approaches for FSRQs is therefore important to verify the reliability of the underlying assumptions of either method, to  estimate 
the associated uncertainties and to constrain the values of the parameters.

The plan of the paper is the following. The selection of the sample and the photometric data we have collected are presented in Section\,\ref{sec:sample}. In Section\,\ref{sec:SED_modelling} we describe the components used to model the spectral energy distributions (SEDs) of our sources and the formalism to obtain the BH mass estimates from the `blue bump' fitting. In Section\,\ref{sect:BH_mass_estimates} we briefly deal with estimates exploiting the SE method, found in the literature, compare them with our estimates, and present the distribution of BH masses obtained from the `blue bump' fitting. Our main conclusions are summarized and briefly discussed in Section\,\ref{sect:conclusions}.

We adopt a standard flat $\Lambda$CDM cosmology with matter density $\Omega_m=0.27$ and Hubble constant $H_0=71\,\hbox{km}\,\hbox{s}^{-1}\,\hbox{Mpc}^{-1}$ \citep{Hinshaw2009}.

\section{The sample}\label{sec:sample}
The Wilkinson Microwave Anisotropy Probe (WMAP) satellite has provided the first all-sky survey at high radio frequencies ($\ge 23\,$GHz). At these frequencies blazars are the dominant radio-source population. We have selected a complete blazar sample, flux-limited at 23 GHz (K band), drawn from the 7-year WMAP point source catalog \citep{Gold2011}.

The basic steps of our selection procedure are the following. We adopted a flux density limit of $S_K=1\,$Jy, corresponding to the WMAP completeness limit \citep{PlanckCollaborationXIII}, and cross-matched the selected sources with the most recent version of the blazar catalogue BZCAT \citep{Massaro2011BZCAT}\footnote{\url{www.asdc.asi.it/bzcat/}}. This search yielded 248 catalogued blazars. To check whether there are additional bona-fide blazars among the other WMAP sources brighter than the adopted flux density limit we have collected data on them from the NASA/IPAC Extragalactic Database (NED)\footnote{\url{ned.ipac.caltech.edu/}}, from the database by \cite{Trushkin2003} and from the catalog of the Australia Telescope Compact Array (ATCA) 20 GHz survey \citep[AT20G,][]{Hancock2011}. Sources qualify as bona-fide blazars if they have: i) a flat radio spectrum ($F_\nu\propto\nu^{-\alpha}$ with $\alpha\le 0.5$); ii) high variability; iii) compact radio morphology. Based on these criteria we have added to our blazar sample 7 sources that satisfy the first two criteria.
The third criterion is satisfied by three of them, whereas for the others no radio image is available in the NED. Our initial sample then consists of 255 blazars, 243 of which have redshift measurements.

Since we are interested in characterizing the optical/UV bump attributed to the accretion disk we have restricted
the sample to the 103 blazars within the area covered by the Eighth Data Release (DR8)\footnote{\url{www.sdss3.org/dr8/}} of the Sloan Digital Sky Survey (SDSS), totalling over 14,000 square degrees of sky and providing simultaneous 5-band photometry with  95\% completeness limiting AB
magnitudes $\textsf{u}$, $\textsf{g}$, $\textsf{r}$, $\textsf{i}$, $\textsf{z} = 22.0$, 22.2, 22.2, 21.3, 20.5, respectively \citep{Abazajian2004}. With the exception of WMAP7\,\#\,274 these objects are in the BZCAT. Moreover, since BL Lacs generally do not show the UV bump, we have dropped from our sample the 19 sources classified as BL Lacs, as well as the 5 sources classified as blazars of uncertain type, keeping only sources classified as FSRQs. The final sample comprises 79 objects, all having spectroscopic redshift measurements.

\subsection{Photometric data}

For these 79 objects we have collected, updated and complemented the photometric data available on the NED, as described in the following.

\subsubsection{SDSS DR8 data}\label{par:SDSSdata}

SDSS counterparts of our FSRQs were searched adopting their low frequency radio coordinates which have uncertainties of $\simeq 1\,$arcsec. Since the SDSS positional uncertainty adds very little to the error (the SDSS positional accuracy is of $\simeq 0.1\,$arcsec) we have chosen a search radius of 3\,arcsec. By construction, all our FSRQs have at least one SDSS counterpart within the search radius. In most cases 
we found a unique counterpart, with the SDSS photometry being consistent with extrapolations
from data at nearby frequencies compliant with a type 1 QSO SED plus the jet emission. Only in eight cases (WMAP7 sources with id numbers 26, 30, 153, 182, 198, 250, 317, 353) we found multiple counterparts.
However in each case one of the sources within the search radius was much (at least 2 magnitudes) brighter than the others,
and had flux densities consistent with those of the FSRQ at nearby frequencies. Thus an unambiguous SDSS counterpart was found for all our FSRQs.
They have a median and an average de-reddened
AB \textsf{r}-band magnitude of 17.67 and   17.63 mag, with a rms dispersion of 1.31 mag.
Thus they are generally much brighter than the 95$\%$ SDSS magnitude limit. Only one FSRQ in the sample, WMAP7 \#\,314, has
an \textsf{r}-band magnitude slightly fainter than that limit.

We have adopted the SDSS magnitudes corrected for Galactic extinction,
as listed in the DR8 catalog and denoted e.g. as \textsf{dered\_g}.
As suggested in the DR8 tutorial\footnote{\url{www.sdss3.org/dr8/algorithms/fluxcal.php\#SDSStoAB}} we have decreased the DR8
\textsf{u}-band magnitudes by 0.04 to bring them to the AB system. The corrections to the magnitudes in the other bands are negligible. In principle some additional extinction may take place within the host galaxy, but we expect it to be negligible because the jet sweeps out any intervening material along its trajectory. The correction for absorption in the intergalactic medium (IGM) is described in sub-section\,\ref{par:IGMext}.  For the redshift range spanned by our sources it may be
relevant only in the \textsf{u} and \textsf{g} bands.

The choice of the effective wavelength corresponding to each  SDSS filter depends on the convolution of the filter spectral response function with the spectral shape of the source. We adopt the effective wavelengths reported in the SDSS tutorial\footnote{\url{skyserver.sdss.org/dr1/en/proj/advanced/color/sdssfilters.asp}}: 3543, 4770, 6231, 7625 and 9134\,{\AA},
for the \textsf{u}, \textsf{g}, \textsf{r},  \textsf{i} and \textsf{z} filters, respectively.

\subsubsection{GALEX data}\label{par:GALEXdata}

We have looked for UV photometry for our FSRQs in the sixth data release,
GR6\footnote{\url{galex.stsci.edu/GR6/}}, of the Galaxy Evolution Explorer (GALEX) satellite \citep{Morrissey2007}.
GALEX provides near-UV (NUV, 1750--2800\,{\AA}) and far-UV (FUV, 1350--1750\,{\AA}) images down to a magnitude limit
$\hbox{AB} \sim 20$--21 with an estimated positional uncertainty of $\simeq 0.5\,$arcsec. We adopt 1535 and 2301\,{\AA}
as the effective wavelengths of the FUV and NUV filters, respectively.

Again the low frequency radio positions of FSRQs were used and a search radius of 3.5 arcsec was adopted.
At least one counterpart was found for 65 objects.
Multiple counterparts were found to correspond to GALEX measurements at different epochs of
the same source (i.e. differences in coordinates were within the positional errors).
Such multi-epoch measurements were found for 24 of our sources.
In these cases we have adopted their weighted average. Whenever the $\hbox{S/N}< 3$,
we have adopted upper limits equal to 3 times the error.

The UV fluxes are very sensitive to extinction within our Galaxy and, in the case of high-$z$ objects, to photoelectric absorption in the intergalactic medium. To correct for Galactic extinction we have used the values of $E(B-V)$ given in the GR6 catalog for each source and the extinction curve by \citet{Cardelli1989}, as updated by \cite{ODonnell1994}, normalized to $A(V)= 3.1\,E(B-V)$. The correction for absorption in the IGM is described in the next section.

\subsubsection{Absorption in the intergalactic medium}\label{par:IGMext}

\begin{figure}
\begin{center}
\includegraphics[width=0.4\textwidth]{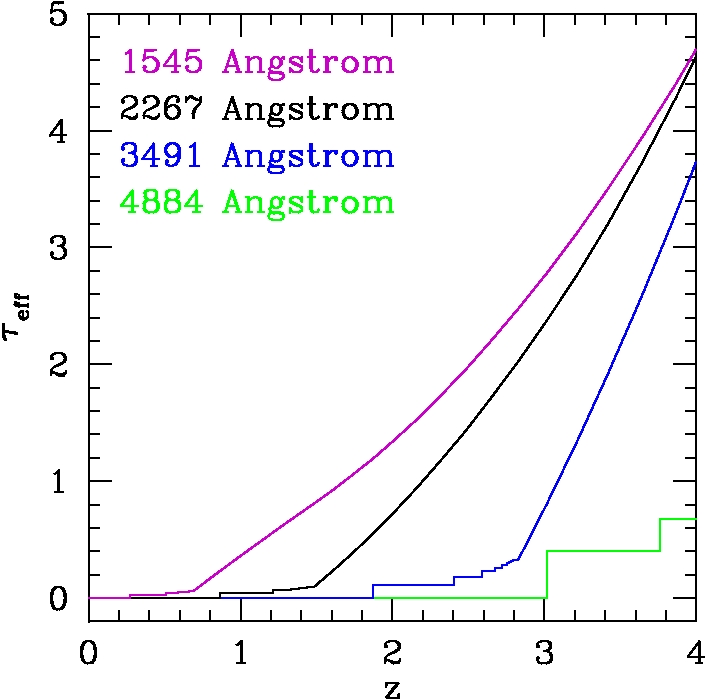}
\end{center}
\caption{Redshift dependent effective optical depth for IGM absorption, averaged over all lines of sight,
 at the effective wavelengths of SDSS \textsf{g} and \textsf{u} bands and of the GALEX NUV and FUV bands.}
\label{fig:taueff_lyman}
\end{figure}

Since we do not know the IGM attenuation along each line of sight we have used the effective optical depth $\tau_{\rm eff}(z)=-\ln[\langle \exp(-\tau)\rangle]$, averaged over all possible lines of sight. We have computed, following \citet{HaardtMadau2012}, $\tau_{\rm eff}(z)$ at
 the effective wavelengths of SDSS \textsf{u} and \textsf{g} filters (the effective optical depth in the other 3 SDSS filters vanishes for the redshift range of interest here) and of the 2 GALEX filters. The results are shown in Fig.~\ref{fig:taueff_lyman} and listed in Table~\ref{tab:optical_depth}. The step-like features are due to Lyman series absorption. We have verified that adopting the spectral response of each filter instead of considering the single effective wavelengths results in a small correction in the flux densities, and in the smoothing of all the edges in the optical depth as a function of redshift. Details on these calculations will be presented in a forthcoming paper 
(Madau \& Haardt, in preparation).


%
\begin{table}
\caption{Redshift dependent effective optical depth for
IGM absorption, averaged over all lines of sight, at the
effective wavelengths of the GALEX NUV and FUV bands and of SDSS the \textsf{g} and \textsf{u} bands.}\label{tab:optical_depth}
\centering
\begin{tabular}{ccccc}
\hline\hline
$z$ & $\tau_{\rm eff}(1545\AA)$ & $\tau_{\rm eff}(2267\AA)$ & $\tau_{\rm eff}(3491\AA)$ & $\tau_{\rm eff}(4884\AA)$ \\
\hline
     0.271    &     0.000    &     0.000    &     0.000    &     0.000    \\
     0.333    &     0.030    &     0.000    &     0.000    &     0.000    \\
     0.399    &     0.030    &     0.000    &     0.000    &     0.000    \\
     0.468    &     0.030    &     0.000    &     0.000    &     0.000    \\
     0.540    &     0.042    &     0.000    &     0.000    &     0.000    \\
     0.615    &     0.049    &     0.000    &     0.000    &     0.000    \\
     0.695    &     0.063    &     0.000    &     0.000    &     0.000    \\
     0.778    &     0.146    &     0.000    &     0.000    &     0.000    \\
     0.865    &     0.232    &     0.000    &     0.000    &     0.000    \\
     0.957    &     0.320    &     0.047    &     0.000    &     0.000    \\
     1.053    &     0.412    &     0.047    &     0.000    &     0.000    \\
     1.154    &     0.506    &     0.047    &     0.000    &     0.000    \\
     1.370    &     0.706    &     0.076    &     0.000    &     0.000    \\
     1.487    &     0.812    &     0.098    &     0.000    &     0.000    \\
     1.609    &     0.926    &     0.233    &     0.000    &     0.000    \\
     1.737    &     1.052    &     0.384    &     0.000    &     0.000    \\
     1.871    &     1.193    &     0.551    &     0.000    &     0.000    \\
     2.013    &     1.352    &     0.737    &     0.112    &     0.000    \\
     2.160    &     1.530    &     0.943    &     0.112    &     0.000    \\
     2.316    &     1.729    &     1.173    &     0.112    &     0.000    \\
     2.479    &     1.951    &     1.428    &     0.184    &     0.000    \\
     2.649    &     2.199    &     1.712    &     0.230    &     0.000    \\
     2.829    &     2.475    &     2.026    &     0.330    &     0.000    \\
     3.017    &     2.782    &     2.376    &     0.808    &     0.000    \\
     3.214    &     3.124    &     2.764    &     1.338    &     0.404    \\
     3.421    &     3.506    &     3.207    &     1.924    &     0.404    \\
     3.638    &     3.930    &     3.713    &     2.573    &     0.404    \\
     3.866    &     4.402    &     4.280    &     3.291    &     0.673    \\
     4.105    &     4.927    &     4.915    &     4.086    &     0.842    \\
     4.356    &     5.511    &     5.626    &     4.964    &     1.195    \\
     4.619    &     6.158    &     6.424    &     5.936    &     2.581    \\
     4.895    &     6.875    &     7.317    &     7.010    &     4.110    \\
     5.184    &     7.669    &     8.318    &     8.196    &     5.794    \\
\hline
\end{tabular}
\end{table}
%

\subsubsection{X-ray data}

We have found ROSAT data for 18 of the 79 FSRQs in our sample. However an inspection of the global SEDs indicates, for all of them,
that X-ray data are clearly far from the fit of the blue bump in terms of a \citet{ShakuraSunyaev1973} accretion disk, adopted in this paper, and more likely
related to other components such as the synchrotron or the inverse Compton ones or the emission from a bright hot X-ray corona.

\subsubsection{WISE data}

We have cross-correlated our FSRQs with the latest version of the Wide-field Infrared Survey Explorer
\citep[WISE;][]{Wright2010} catalog\footnote{\url{wise2.ipac.caltech.edu/docs/release/allsky/}}. Again,
the coordinates of low radio frequency counterparts were adopted and a search radius of 6.5\,arcsec was chosen,
 consistent with  WISE positional uncertainty\footnote{According to the WISE Explanatory 
Supplement \citep{Cutri2013}, sources with $S/N \sim20$ have a typical
 rms positional uncertainty of 0.43\,arcsec. Our sources generally have a
 much lower $S/N$ ratio and the astrometric uncertainty scales as $(S/N)^{-1}$ \citep[e.g.][]{Condon1998,Ivison2007}.
 For the typical $S/N$ values of our sources, $S/N=3$--5, the search radius correspond to positional errors in the range 2.3--$3.8\sigma$.}. We found WISE counterparts with $\hbox{S/N}\geq3$
in at least one band for all the 79 FSRQ in the sample. In the WISE bands where $\hbox{S/N}<3$  we have adopted an upper limit equal to 3 times the error. Multiple WISE sources were found within the search radius for the FSRQs WMAP7 \#\,30, 126, 278, 317 and 378.  In these cases we have chosen the brightest WISE source as the most likely counterpart. In all cases, the other sources were at least 2 magnitudes dimmer.

\subsection{{\it Planck} data}
In the {\it Planck} Early Release Compact Source Catalog \citep[ERCSC;][]{PlanckCollaborationVII2011} we have found counterparts for 72 out of our 79 FSRQs; 47, 39, and 68 of them have ERCSC flux densities at 70, 44, and 30~GHz, respectively and 63 have ERCSC flux densities at frequencies $\ge 100\,$GHz.

\begin{figure}
\begin{center}
\includegraphics[width=0.4\textwidth]{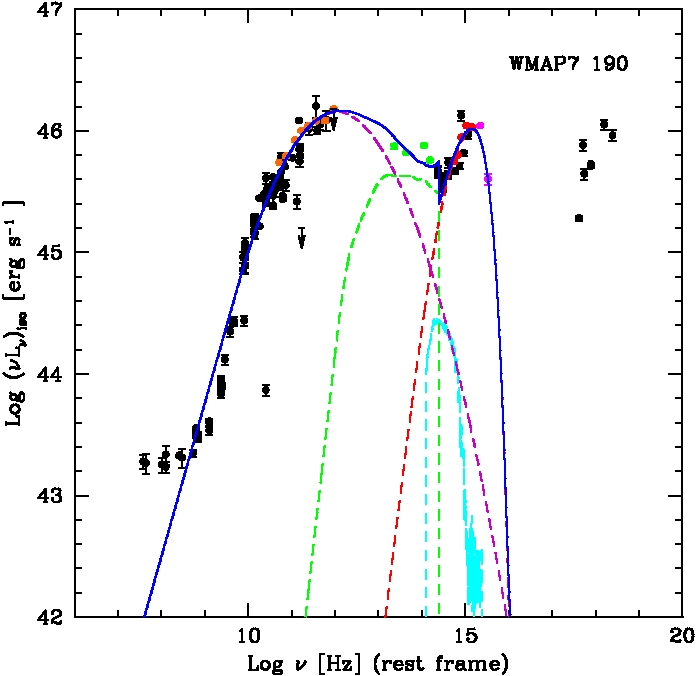}
\end{center}
\caption{Example of a SED fit (WMAP7\,\#\,190). Solid blue parabola: total SED, which includes synchrotron emission, host galaxy, disk and torus emissions; dashed violet line: synchrotron from the jet; green dashed line: torus; dashed cyan line: host galaxy, taken to be a passive elliptical with $M_{R}=-23.7$; dashed red line: accretion disk. Orange points: Planck data; green: WISE data; red: SDSS data; magenta: GALEX data.
 Black points: data taken from the NASA/IPAC Extragalactic Database (NED). Note that, at variance with
 what  was done to compute $L_d$ (see text), the luminosities shown here are computed assuming isotropic emission. }
\label{fig:SED_190}
\end{figure}

\section{SED modeling}\label{sec:SED_modelling}

Of the 79 FSRQs in our sample, 54 (i.e. 68\%) show clear evidence of the optical/UV bump, interpreted as the emission from
a standard optically thick, geometrically thin accretion disk model \citep{ShakuraSunyaev1973}.
 As illustrated by the example shown in Fig.~\ref{fig:SED_190}, the global SEDs are modeled
taking into account several
additional components: the Doppler boosted synchrotron continuum 
modeled following \citet{Donato2001}; 
a passive elliptical host galaxy template  \citep[see, e.g.,][]{Giommi2012}; the dusty AGN torus 
emission based on a type-1 QSO template (BQSO1) from the 
\citet{Polletta2007}\footnote{\url{www.iasf-milano.inaf.it/$\sim$polletta/templates/swire\_templates .html}} SWIRE template library.

The fit of the global SED was made using 6 free parameters. Four are those of the blazar sequence model for 
the synchrotron emission \citep[the 5~GHz luminosity, the 5~GHz spectral index, the junction frequency between the low-  
and the high-frequency synchrotron 
template and the peak frequency of $\nu{\rm L}_\nu$,][]{Donato2001}. The remaining two parameters refer to the accretion disk model
 (i.e. the normalization and the peak frequency). The other components are fixed. The host galaxy template is an 
elliptical \citep{Mannucci2001} with an absolute magnitude of $M_{R}=-23.7$, as in \citet{Giommi2012}. 
The normalization of the torus template was computed from that of the accretion disk emission, requiring 
that the torus/accretion disk luminosity ratio is equal to that of the \citet{Polletta2007} BQSO1 template. 

We stress that accurate fits of the global SEDs are beyond the scope of the present paper whose
main purpose is the  estimate the black-hole masses by fitting the optical/UV bump with a \citet{ShakuraSunyaev1973} model,
as discussed in the following. The consideration of the other components,
fitted taking into account all the data we have collected, is mainly relevant to check
whether they may contaminate the emission from the accretion disk. In many cases, the
lack of simultaneity of the measurements does not allow reliable fits of the other components.
Still for the 54 objects with clear evidence of the blue bump the data where enough either to
estimate the amount of contamination or to signal points that should be better
taken as upper limits to the blue bump emission.

The thermal emission from the accretion disk is modeled as a combination of black-bodies with temperatures
depending on the distance, $R$, from the black-hole \citep[see, e.g.,][]{Frank2002}.
The flux density observed at a frequency $\nu_o$ is given by:
\begin{equation}
 \label{eq:SS_discflux}
 F_\nu(\nu_o) = \nu_o^3\frac{4\pi h_P \cos(i) }{c^2 {D_A}^2}\int_{R_\star}^{R_{\rm out}}\frac{R\,dR}{e^{h_P(1+z) \nu_o/kT(R)}-1},
\end{equation}
where $D_A$ is the angular diameter distance to  the blazar, $k$ is the Boltzmann constant, $z$ is the redshift of the source, $h_P$ is the Planck constant, $c$ is the speed of light, $R_\star$ and $R_{\rm out}$ are the inner and outer disk radii, respectively. The radial temperature profile, $T(R)$, is given by:
\begin{equation}
 \label{eq:SS_temp}
 T^4(R)= \frac{3GM_\bullet\dot{M}}{8\pi R^3\sigma}\left(1-\sqrt{\frac{R_\star}{R}} \right),
\end{equation}
where $G$ is the gravitational constant, $M_\bullet$ is the black hole mass, $\dot{M}$ is its accretion rate, $\sigma$ is the Stefan-Boltzmann constant. $R_\star$ is the radius of the last stable orbit that, for a Schwarzschild black hole, is  $R_\star=3R_S$,  $R_S = 2GM_\bullet/c^2$ being the Schwarzschild radius. The results are insensitive to the chosen value for $R_{\rm out}$ provided that $R_{\rm out}\gg R_\star$; we choose $R_{\rm out}=100R_S$.

Since the emission of the disk is anisotropic (the flux density measured by an observer is proportional to $\cos i$), the monochromatic luminosity is related to the flux density by \citep{Calderone2012}
\begin{equation}\label{eq:disc_lum}
\nu_e L_\nu(\nu_e) = {2\pi D_L^2\nu_o F_\nu(\nu_o)\over \cos i},
\end{equation}
where $\nu_e=(1+z)\nu_o$ is the frequency at the emission redshift, $z$, and $D_L(z)$ is the luminosity distance.

For the 54 FSRQs showing the optical/UV bump the fit of the accretion disk model to it was done using only the SDSS (available for all of them) and the GALEX data
 (available for all but 7 of them). Using the standard minimum $\chi^2$ technique we have obtained the values 
of the two free parameters, the normalization and the peak frequency, $\nu_{\rm peak}$ (in terms of $\nu L_\nu$). The total disk luminosity, $L_d$, can then be computed integrating eq.~(\ref{eq:disc_lum}) over frequency. The derived values of $\nu_{\rm peak} L_\nu(\nu_{\rm peak})$,  $L_d$ and $\nu_{\rm peak}$ are given in Table~\ref{tab:sample_prop}. The accretion rate is  $\dot{M} = {L_d}/({\eta c^2})$ where $\eta$ is the mass to light conversion efficiency for which we adopt the standard value $\eta=0.1$.

\begin{table*}
\caption{Best fit values of the big blue bump parameters.}\label{tab:sample_prop}
\centering
\begin{tabular}{ccccc}
\hline\hline

WMAP ID & $\log\left({\nu_{\rm peak}\over \rm Hz}\right)$ & $\log\left({\nu_{\rm peak} L_\nu(\nu_{\rm peak})\over \rm erg\,s^{-1}}\right)$ & $\log\left({L_d\over \rm erg\,s^{-1}}\right)$ & $\log\left({M_\bullet\over M_\odot}\right)$ \\
\hline

      9    &     15.52    &     45.62    &     45.90    &      8.53    \\
     26    &     15.72    &     46.35    &     46.63    &      8.49    \\
     27    &     15.29    &     46.00    &     46.26    &      9.18    \\
     31    &     15.42    &     46.36    &     46.63    &      9.09    \\
     39    &     15.32    &     45.37    &     45.64    &      8.80    \\
     42    &     15.32    &     45.94    &     46.21    &      9.09    \\
     89    &     15.42    &     45.32    &     45.59    &      8.57    \\
    137    &     15.62    &     46.78    &     47.06    &      8.91    \\
    150    &     15.33    &     45.81    &     46.07    &      9.01    \\
    153    &     15.62    &     46.24    &     46.52    &      8.64    \\
    155    &     15.39    &     45.81    &     46.08    &      8.88    \\
    160    &     15.32    &     46.34    &     46.61    &      9.28    \\
    166    &     15.32    &     45.69    &     45.96    &      8.96    \\
    169    &     15.42    &     45.92    &     46.19    &      8.87    \\
    173    &     15.42    &     46.49    &     46.77    &      9.17    \\
    179    &     15.42    &     45.42    &     45.69    &      8.62    \\
    182    &     15.57    &     46.05    &     46.33    &      8.64    \\
    186    &     15.17    &     46.42    &     46.67    &      9.61    \\
    190    &     15.15    &     45.72    &     45.96    &      9.30    \\
    191    &     15.52    &     46.18    &     46.45    &      8.80    \\
    195    &     15.39    &     46.10    &     46.37    &      9.04    \\
    198    &     15.32    &     45.37    &     45.64    &      8.80    \\
    203    &     15.52    &     45.21    &     45.48    &      8.32    \\
    208    &     15.42    &     46.29    &     46.56    &      9.06    \\
    221    &     15.52    &     46.28    &     46.56    &      8.86    \\
    224    &     15.36    &     45.93    &     46.20    &      8.99    \\
    228    &     15.32    &     45.40    &     45.67    &      8.81    \\
    232    &     15.52    &     45.12    &     45.39    &      8.27    \\
    236    &     15.32    &     44.95    &     45.21    &      8.58    \\
    250    &     15.35    &     45.97    &     46.24    &      9.04    \\
    265    &     15.67    &     45.72    &     46.00    &      8.28    \\
    278    &     15.17    &     45.97    &     46.23    &      9.39    \\
    284    &     15.62    &     45.70    &     45.97    &      8.36    \\
    295    &     15.92    &     46.24    &     46.52    &      8.04    \\
    306    &     15.62    &     45.44    &     45.71    &      8.23    \\
    307    &     15.22    &     45.78    &     46.04    &      9.20    \\
    310    &     15.42    &     45.75    &     46.02    &      8.79    \\
    311    &     15.31    &     46.21    &     46.47    &      9.25    \\
    316    &     15.38    &     46.36    &     46.63    &      9.19    \\
    317    &     15.42    &     44.78    &     45.05    &      8.30    \\
    327    &     15.71    &     46.46    &     46.74    &      8.58    \\
    402    &     15.42    &     45.92    &     46.19    &      8.87    \\
    407    &     15.82    &     46.74    &     47.02    &      8.49    \\
    412    &     15.40    &     46.87    &     47.14    &      9.40    \\
    415    &     15.37    &     45.24    &     45.51    &      8.63    \\
    417    &     15.32    &     46.55    &     46.82    &      9.39    \\
    428    &     15.39    &     46.34    &     46.61    &      9.14    \\
    430    &     15.67    &     46.66    &     46.93    &      8.75    \\
    434    &     15.42    &     45.94    &     46.21    &      8.88    \\
    452    &     15.38    &     46.32    &     46.59    &      9.16    \\
    455    &     15.32    &     45.87    &     46.14    &      9.05    \\
    458    &     15.52    &     45.21    &     45.49    &      8.32    \\
    462    &     15.37    &     46.11    &     46.38    &      9.07    \\
    470    &     15.52    &     46.03    &     46.30    &      8.73    \\

\hline
\end{tabular}
\end{table*}
%
An analysis of eq.~(\ref{eq:SS_discflux}) indicates that the main contribution to the integral comes from a region around the radius $R_{\rm peak}=(49/36)R_\star$ where the temperature profile $T(R)$ [eq.~(\ref{eq:SS_temp})] reaches its maximum value $T_{\rm max}$. The integral  over $R$, to compute $L_d$ (hence $\dot{M}$), can then be approximately evaluated with the steepest descent method. The calculation was made setting $i=0$. Then, introducing the value of $T_{\rm max}=T(R_{\rm peak})$ in the Wien's displacement law,  $\nu_{\rm peak}/T_{\rm max}\simeq 5.879\times 10^{10}\,\hbox{Hz}\,\hbox{K}^{-1}$, we get an estimate of the black hole mass: $M_\bullet/ 10^9 M_\odot \simeq 0.46 ({\nu_{\rm peak}/3\times 10^{15}\,\hbox{Hz}})^{-2} ({\dot{M}}/{M_\odot\,{\rm yr}^{-1}})^{1/2}$. This result shows that the estimate of $M_\bullet$ is quite sensitive to the value of $\nu_{\rm peak}$. One may then wonder whether associating it to $T_{\rm max}$ is a sufficiently good approximation. To answer this question we have computed $M_\bullet$ by numerically solving the equation $d\log(\nu_e L_\nu(\nu_e))/d\log(\nu_e)=0$ for all values of $\nu_{\rm peak}$ and $L_{d}$ found for our sources. Remarkably, we find that the exact values of $M_\bullet$ strictly follow the dependencies on $\dot{M}$ and $\nu_{\rm peak}$ given by the approximate solution, with a coefficient lower by a factor 0.76. The black hole masses implied by the \citet{ShakuraSunyaev1973} model can then be accurately computed using the simple equation:
%
\begin{equation}
 \label{eq:M_BH}
 {M_\bullet\over 10^9 M_\odot } \simeq 0.35 \left({\nu_{\rm peak}\over 3\times 10^{15}\,\hbox{Hz}}\right)^{-2}  \left( \frac{\dot{M}}{M_\odot \;{\rm yr}^{-1}}\right)^{1/2}.
\end{equation}
%
The results for our FSRQs are reported in Table~\ref{tab:sample_prop}.

The statistical errors associated with $\log(\nu_{\rm peak})$ and $\dot{M}$ were computed
utilizing the standard criteria based on the $\chi^2$ statistics \citep[e.g.,][]{Cash1976},
with errors estimated adding in quadrature the measurement uncertainties and the estimated spread
of data points due to variability. The uncertainty on $\log(\nu_{\rm peak})$ was found to be in the
 range 0.02--0.09, that on $\log(\dot{M})$ in the range 0.02--0.10, depending on the data quality.
The errors on $\log(M_\bullet)$ cannot be obtained by simply summing the two contributions in
quadrature because $\log(\nu_{\rm peak})$ and $\dot{M}$ are not independent. From the distribution
 of $\log(M_\bullet)$ obtained varying the two quantities within their 68\% confidence interval,
 we found uncertainties in the range 0.1--0.3.

The uncertainties on the IGM absorption correction due to variations of the effective optical depth
with the line of sight are unknown. An insufficient correction for UV absorption leads to an underestimation
of $\nu_{\rm peak}$ and to an overestimation of $M_\bullet$, while an overcorrection has the opposite effect.
 However, since all of our FSRQs but one (namely WMAP7 \#\,137 that has a redshift $z=3.4$) are at $z<2.5$ the
corrections for IGM absorption are relatively small. Ignoring such correction would lead to a mean overestimate of $\log(M_\bullet)$
by 0.04 for the 18 objects with $z<1$, of 0.09 for the 17 objects with $1<z<1.5$ and of 0.11 for the 13 objects at $1.5<z<2$.
For WMAP7 \#\,137 and for the 5 objects at $2<z<2.5$ the variation of $\nu_{\rm peak}$ is compensated by that of $L_d$ so
that the average difference between corrected and uncorrected estimates is negligible.

Further uncertainties are associated with the choice of the model and of its parameters.
As pointed out in Sect.~\ref{par:intro}, the adopted accretion disk model assumes a non-rotating BH,
although the chosen value of the radiation efficiency, $\eta=0.1$, is above the maximum efficiency
for a  Schwarzschild BH. However, \citet{Calderone2012}, using the \citet{Li2005} software package,
 found that the \citet{ShakuraSunyaev1973} model with $R_\star=3R_S$, as used here, mimics quite
 well the SED for an optically thick, geometrically thin accretion disk around a Kerr BH with a
spin parameter $a\simeq 0.7$, corresponding to a maximum radiative efficiency $\eta=0.1$.
For this choice of $\eta$ our BH mass estimates are therefore little affected by having
neglected the general relativistic effects associated with a Kerr BH. Based on the analysis presented
in Appendix A4 of \citet{Calderone2012} we find for a pure Schwarzschild model ($a=0$, $\eta=0.06$)
a BH mass lower by a factor of 0.6 while for a Kerr model with a maximum possible radiative efficiency
 ($a = 0.998$) we find a BH mass higher by a factor of 1.75. Note however that the latter factor is a generous upper limit
since the boundaries of the range of values of $\eta$ for which \citet{Shankar2009} achieved a good match to the overall
 shape of the BH mass function are $0.06\le \eta \le 0.15$. The effect of the choice of the inclination angle $i$ should
 be minor given the model and observational indications that $i \lsim 5^\circ$; even
if we double this value we get $\cos(10^\circ)=0.985$.

Summing up in quadrature the uncertainties listed above we end up with errors on $\log(M_\bullet)$ in the range 0.2--0.4. 
These estimates should be taken as lower limits since they do not include all the uncertainties in the theoretical
 accretion disk model, which are difficult to quantify.

\section{Black hole mass estimates}\label{sect:BH_mass_estimates}

\subsection{Estimates with the single epoch virial method}\label{sect:SEestimates}
Black hole mass estimates obtained with the single epoch virial method (SE method) are available in the literature for several FSRQs in our sample.
\citet{Shaw2012} derived them for a sub-sample of blazars selected from the First Catalog of Active Galactic Nuclei detected by the Fermi Large Area Telescope \citep[1LAC,][]{Abdo2010}, including 24 of our FSRQs. They considered several estimators exploiting continuum and emission line  (H$\beta$, MgII, CIV) measurements. We preferred the estimates based on line measurements
to those obtained by using the continuum luminosity because the latter is liable to contamination from the jet synchrotron emission. More precisely, we choose, in order of preference,  estimates derived from H$\beta$ and MgII lines for the blazars at redshift $z<1$ and the ones derived from the MgII and CIV lines for the blazars at higher redshifts \citep[see][for more details]{Shaw2012}.

\citet{Shen2011} estimated the BH masses for a sample of quasars drawn from the SDSS-DR7 quasar catalog \citep{Schneider2010}, including 36 objects in common with our sample. Seventeen of the latter belong also to the \citet{Shaw2012} sample.
However, although \citet{Shen2011} give measurements of line luminosities and FWHM, the fiducial BH masses reported by them are based 
on continuum rather than line luminosity. Thus we have used their line data to recompute the BH masses for the 36 objects in common 
with our sample. Since several line measurements are present for a given source, following \citet{Shen2011}
we adopted H$\beta$, MgII, and CIV line measurements for the blazars at redshift $z<0.7$, $0.7\leq  z<1.9$,
and $z\geq1.9$, respectively. The average logarithmic ratio of the black hole mass estimates based on line luminosities to the fiducial values given by \citet{Shen2011}, based on continuum luminosities, for the 36 sources is $\langle\log(M_{\bullet,\rm Shen,lines}/M_{\bullet,\rm Shen})\rangle=-0.17$ with a rms dispersion of 0.23. This suggests that indeed the continuum 
luminosities are likely contaminated by the optical emission 
from the jet, as argued by \citet{Shen2011}.
Our re-evaluations of the BH mass estimates of the \citet{Shen2011} blazars are in good agreement with those by
 \citet{Shaw2012} for the 17 blazars in common. The average logarithmic ratio of the two estimates is
$\langle\log(M_{\bullet,\rm Shaw}/M_{\bullet,\rm Shen,lines})\rangle=0.01$, with a dispersion of 0.22.


Since the analysis by \citet{Shaw2012} is focussed on FSRQs, for the comparison with the BH mass estimates obtained from the fitting of the blue bump we preferred their estimates for the objects in common with \citet{Shen2011}.
For the other \citet{Shen2011} blazars in our sample we have adopted our new determinations of BH masses
based on line luminosities.  The corresponding uncertainties were computed applying the standard error propagation, 
taking into account measurement errors on line luminosities and FWHMs as well as the errors on the coefficients of the
relations between these quantities and the BH mass, as reported in \cite{Shen2011} and references therein. The latter 
errors are the main contributors to the global uncertainties.

Black hole mass estimates for one additional object in our sample, WMAP7 \#\,250, were reported by \citet{Kaspi2000}
and \citet{Shang2007}. We adopted the latter, more recent estimate.

\begin{figure}
 \centering
 \includegraphics[width=0.4\textwidth]{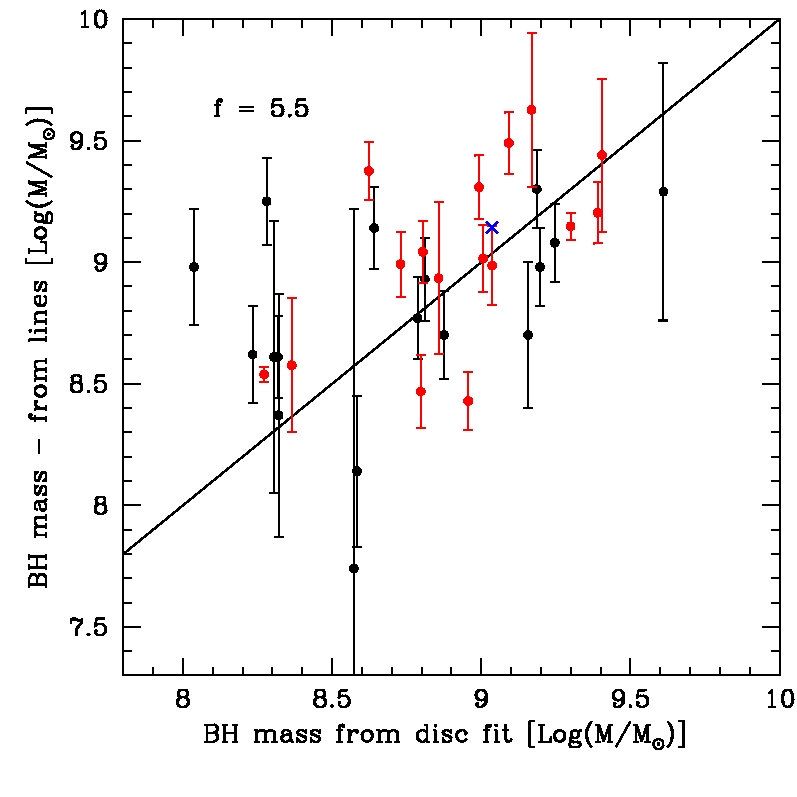}
 \caption{Comparison of the black hole mass estimates. Estimates with the SE method against those
from fitting the accretion disk SED. The black points are from \citet{Shaw2012},
the red points are our estimates using line data from  \citet{Shen2011}, the blue cross refers to the estimate
by \citet{Shang2007} corrected as mentioned in the text. No error was reported for this estimate.}
\label{fig:Fermi_SDSS}
\end{figure}

\subsection{The \it{f}-factor}
The BH masses estimated with the SE method assume that the optical/UV line emission is coming mainly from the BLR, located at a radial distance $R_{\rm BLR}$ from the central black hole. Assuming that the BLR clouds are in virial equilibrium, $M_{\bullet}$ is given by eq.~(\ref{eq:virial_BHmass}). 
There are two commonly used measures of the cloud velocity $\Delta  V$: the line FWHM and the dispersion of its Gaussian fit. 
We adopt the second one, i.e. $\Delta  V=\sigma_{\rm line}$. $R_{\rm BLR}$ is estimated using empirical analytic relations with continuum or line luminosities.  With these assumptions and notation for an isotropic velocity field we have  $f=3$ \citep{Netzer1990}. 
This is however an over-simplified model. In practice, the value of $f$ is empirically determined, 
but there is no consensus on its value \citep[see][and references therein]{Park2012a,Park2012b}.
Values claimed in the literature differ by a factor of 2, from $f\simeq 2.8$ \citep{Graham2011} to $f\simeq 5.5$ \citep{Onken2004}.
Note that for face-on objects (such as FSRQs) the average virial coefficient $f$ may be larger than for optically selected QSOs
 (with random orientations) if the BLR has a flattened geometry \citep{Decarli2011}.
 The empirical relations  used by \citet{Shen2011} and \citet{Shaw2012} implicitly assume $f=5.5$ since this value was used, 
following \citet{Onken2004}, in calibrating the reverberation mapping BH masses which in turn were used as standards 
to calibrate SE mass estimators. \citet{Shang2007} 
followed a different approach, adopting $f =3$. The latter authors also used a slightly different cosmology. 
We have corrected their BH mass estimate to homogenize it with the others. 


\subsection{Comparison of black hole mass estimates}\label{sec:BH_mass_comparison}

Thirty-four of the 54 blazars for which we could derive the BH masses with the blue bump fitting method also have
published estimates with the SE method. In Figure~\ref{fig:Fermi_SDSS} we compare the results from the two methods, after having  homogenized the SE estimates as described above.
They are well correlated: the Spearman test yields a  99.96\% (i.e. $3.5\sigma$) significance of the correlation.
 The SE method with  $f=5.5$ yields, on average, slightly higher values of $M_\bullet$.
We find  an average $\langle\log(M_{\bullet,\rm SE}/M_{\bullet,\rm blue\, bump})\rangle = 0.09$
with a rms dispersion of 0.40 dex. For comparison, the uncertainty of the SE method
is of 0.4--0.5 dex \citep{VestergaardPeterson2006,Park2012b} and
that of the blue bump method is $\gsim 0.2$--0.4 (see Sect.\,\ref{sec:SED_modelling}). Thus the rms difference is fully accounted for by the uncertainties of the two methods.
The offset between the two estimates would be removed setting $f=4.5$, well within the range of current estimates.
However, in view of the large uncertainties, reading this as an estimate of $f$ would constitute an over-interpretation of the data.
On the other hand, the consistency of the two methods strongly suggests that neither is badly off.

\begin{figure}
\begin{center}
\includegraphics[width=0.4\textwidth]{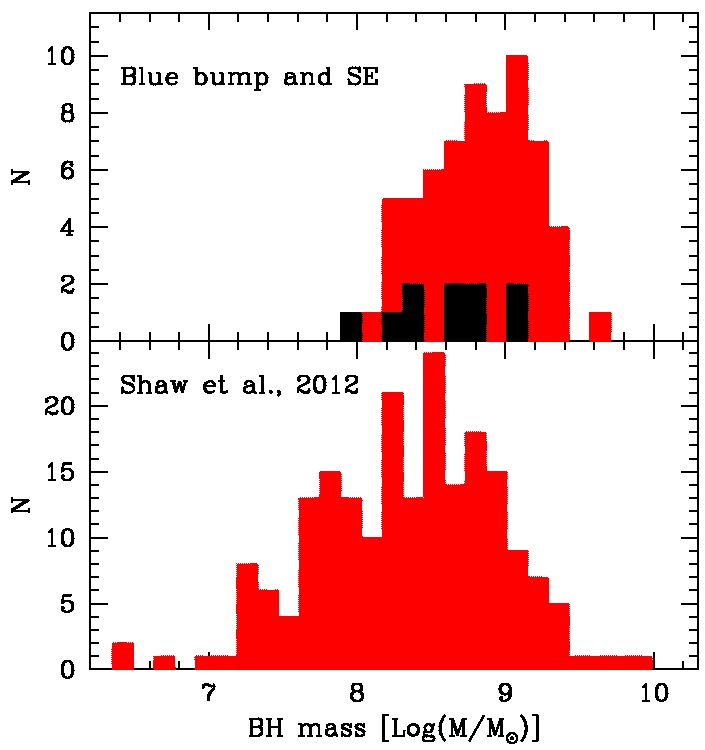}
\end{center}
\caption{Distributions of black hole masses. The upper panel shows in red the distribution for our 54 objects
having estimates via blue bump fitting and, in black, for 10 additional objects in the sample for which we have BH mass estimates via the SE method only,
homogenized as described in the text and decreased by 0.09 dex to remove the mean offset with blue bump estimates.
The lower panel shows the distribution for all 1LAC blazars \citep{Shaw2012} with masses decreased by 0.09 dex.}
\label{fig:BHmass_histogram}
\end{figure}

\subsection{Distribution of black hole masses}\label{sec:BH_mass_distribution}

In the upper panel of Fig.~\ref{fig:BHmass_histogram} we report in red the distribution of BH masses obtained by means of the blue bump fitting for 54 of our FSRQs and, in black, the distribution for the 10 additional ones for which only SE estimates are available.
The estimates for the latter objects have been first homogenized as described above and then decreased by 0.09 dex to
 correct for the mean offset with the blue bump results. The lower panel shows, for comparison,
the distribution for 1LAC blazars in the \citet{Shaw2012} sample, again decreasing the BH masses by 0.09 dex.
Whenever \citet{Shaw2012} provide more than one mass estimate for a single object we made a choice abiding by the order of preference mentioned in Sect.\,\ref{sect:SEestimates}.

The figure shows that our 64 FSRQs are associated with very massive BHs ($M_\bullet \gtrsim10^{7.8}\,M_\odot$) with a median value of $6.8\times10^8\,M_\odot$. The median BH mass changes little (it becomes $7.4\times10^8~M_\odot$) if we restrict ourselves to the 54 FSRQ with BH mass estimates via blue bump fitting. The decline of the distribution at lower masses may be a selection effect: we have selected radio-bright objects ($S_{23\rm GHz}\ge 1\,$Jy) and the 15 (19\%) FSRQs in our sample that do not show a detectable blue bump nor have SE estimates of the BH mass may well 
be associated with lower values of $M_\bullet$.
On the other hand, our results are also consistent with the theoretical and observational studies which suggest that radio loud AGNs are 
generally associated with the most massive black holes \citep[$M_\bullet \gtrsim 10^8\,M_\odot$, e.g.][]{ChiabergeMarconi2011}.  The fast decline of the distribution above $M_\bullet\simeq 10^{9.4}\,M_\odot$, suggesting some upper bound to BH masses, is more likely to be real.

The BH mass distribution of \citet{Shaw2012} blazars adds support to the conclusion that {\it blazar} BH masses either below $M_\bullet \sim 10^{7.4}\,M_\odot$ or above $M_\bullet \sim 10^{9.6}\,M_\odot$ are rare. In this context it is worth noticing that errors in BH mass estimates tend to populate the tails of the distribution by an effect analogous to the Eddington bias: objects preferentially move from highly populated to less populated regions. Thus in particular the highest mass tail may be overpopulated (while the effect on the low mass tail may be swamped by selection effects).

\begin{figure} \centering
\subfigure[]{\includegraphics[width=0.4\textwidth]{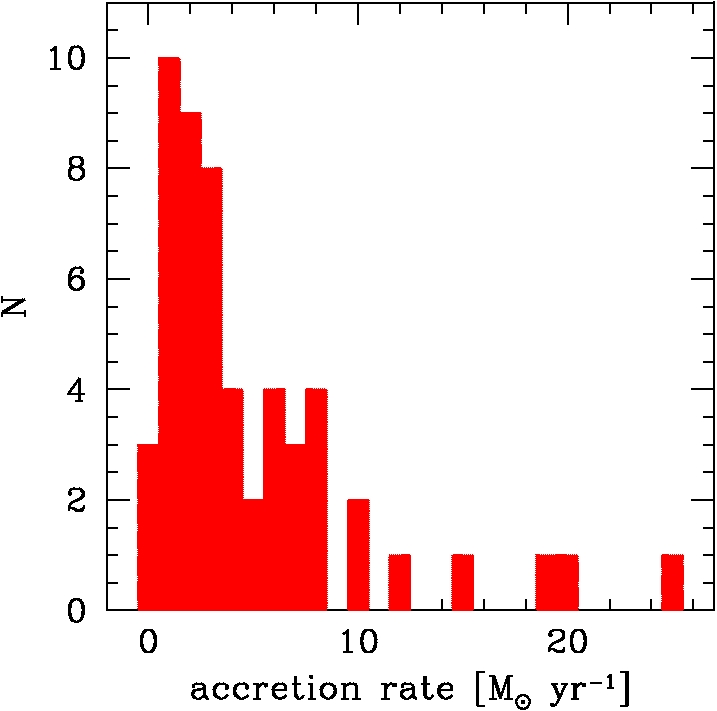}}\qquad
\subfigure[]{\includegraphics[width=0.4\textwidth]{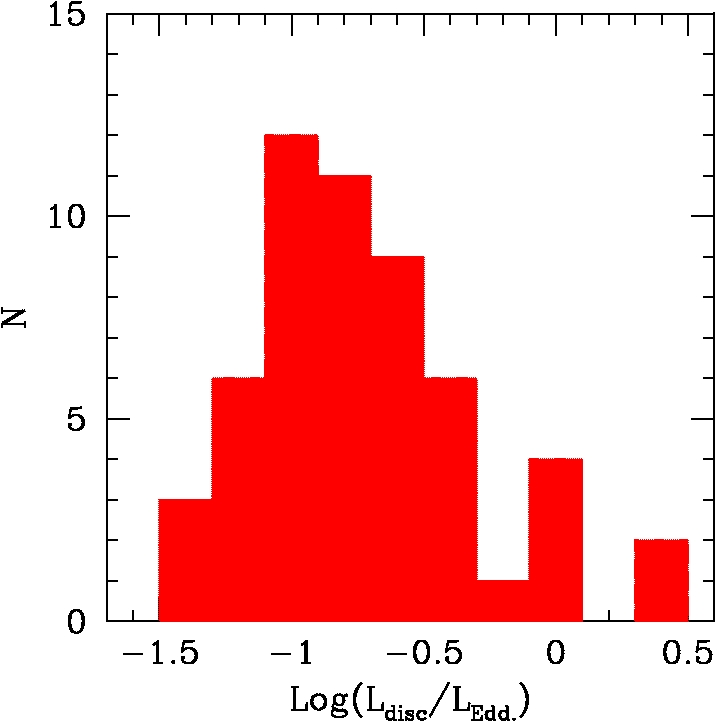}}
\caption{Distributions of the accretion rate (top) and of the Eddington ratio (bottom) for the our FSRQs with evidence of blue bump. }
\label{fig:Edd_distr}
\end{figure}

In Figure~\ref{fig:Edd_distr} we report the distributions of the accretion rates (top panel) and of the Eddington ratios
(bottom panel) for the 54 FSRQ in the sample for which we have estimated the BH mass fitting the blue bump of the spectrum,
as reported in Table~\ref{tab:sample_prop}. We find a median Eddington ratio of 0.16 and a median accretion rate of $2.8\,M_\odot\,\hbox{yr}^{-1}$. The few extreme values of these parameters must be taken with special caution on account of
the uncertainties affecting our estimates.

The BH mass turns out to be anti-correlated with the disk peak frequency. The Spearman's test gives a probability of the null hypothesis (no correlation) $p=6.8\times 10^{-5}$. The anticorrelation follows from eq.~(\ref{eq:M_BH}) due to the weak dependence of $M_\bullet$ on the accretion rate and the limited range of $\dot{M}$ spanned by our blazars.



\section{Discussion and conclusions}\label{sect:conclusions}

We have compared black hole mass estimates based on fitting the blue
bump with a \citet{ShakuraSunyaev1973} model with those obtained with the commonly used single epoch virial method
(SE method) for a complete sample of FSRQs drawn from the WMAP 7-yr catalog, all
with measured spectroscopic redshift. The sample comprises 79 objects with $S_{23\rm GHz}\ge 1\,$Jy, 54 of which (68\%) have a
clearly detected `blue bump' from which the black hole mass could be inferred. FSRQs are the
 AGN population best suited for such a comparison because there is good evidence that their jets are highly aligned with the line-of-sight,
 suggesting that the accretion disk should be almost face-on, thus minimizing the uncertainty on the inclination angle that
bewilders black hole mass estimates for the other AGN populations.

The mass estimates obtained from the two methods are well correlated. If the calibration factor $f$ of the SE relation,
eq.~(\ref{eq:virial_BHmass}), is set $f=4.5$, well within the range of recent estimates,
 the mean logarithmic ratio of the two mass estimates is
$\langle\log(M_{\bullet,\rm SE}/M_{\bullet,\rm blue\, bump})\rangle =0$ and its dispersion is 0.40, close to that expected from uncertainties of the two methods. The fact that the two independent methods agree so closely in spite of all the potentially large 
uncertainties associated with each (see Sects.\,\ref{par:intro} and \ref{sec:SED_modelling}) lends strong support to both of them. However the agreement is only statistical, and individual estimates of black hole masses must be taken with caution.

The distribution of black-hole masses for the 54 FSRQs in our sample with a well detected blue bump has a median value of $7.4\times10^8\,M_\odot$. It declines at the low mass end, consistent with other indications that radio loud AGNs are
generally associated with the most massive black holes, although the decline may be, at least partly, due to the source selection. The distribution drops above $M_\bullet = 2.5\times 10^9\,M_\odot$, implying
 that ultra-massive black holes associated with FSRQs must be rare.

\begin{acknowledgements}
We gratefully acknowledge very useful comments from an anonymous referee. Work supported in part by ASI/INAF agreement n. I/072/09/0 and by INAF through the PRIN 2009 ``New light on the early Universe with sub-mm spectroscopy''.
This publications has made use of data products from the Wide-field Infrared Survey Explorer, which is a joint project of the
University of California, Los Angeles, and the Jet Propulsion Laboratory/California Institute of Technology, funded
by the National Aeronautics and Space Administration, and of the NASA/IPAC Extragalactic Database (NED)
which is operated by the Jet Propulsion Laboratory, California Institute of Technology, under contract
with the National Aeronautics and Space Administration.
\end{acknowledgements}

{70}

\begin{thebibliography}{70}

\bibitem[Abazajian et al.(2004)]{Abazajian2004} Abazajian, K., Adelman-McCarthy, J.~K., Ag{\"u}eros, M.~A., et al.\ 2004, \aj, 128, 502

\bibitem[Abdo et al.(2010)]{Abdo2010} Abdo, A.~A., Ackermann, M., Ajello, M., et al.\ 2010, \apj, 715, 429

\bibitem[Ajello et al.(2012)]{Ajello2012} Ajello, M., Shaw, M.~S., Romani, R.~W., et al.\ 2012, \apj, 751, 108


\bibitem[Assef et al.(2012)]{Assef2012} Assef, R.~J., Frank, S., Grier, C.~J., et al.\ 2012, \apjl, 753, L2

\bibitem[Blandford \& McKee(1982)]{BlandfordMcKee1982} Blandford, R.~D., \& McKee, C.~F.\ 1982, \apj, 255, 419
\bibitem[Bentz et al.(2009)]{Bentz2009a} Bentz, M.~C., Peterson, B.~M., Pogge, R.~W., \& Vestergaard, M.\ 2009, \apjl, 694, L166
\bibitem[Calderone et al.(2013)]{Calderone2012} Calderone G., Ghisellini G., Colpi M., Dotti M., 2013, MNRAS, 431, 210


\bibitem[Cardelli et al.(1989)]{Cardelli1989} Cardelli, J.~A., Clayton, G.~C., \& Mathis, J.~S.\ 1989, \apj, 345, 245


\bibitem[Cash(1976)]{Cash1976} Cash, W.\ 1976, \aap, 52, 307



\bibitem[Chiaberge \& Marconi(2011)]{ChiabergeMarconi2011} Chiaberge, M., \& Marconi, A.\ 2011, \mnras, 416, 917

\bibitem[Condon et al.(1998)]{Condon1998} Condon, J.~J., Cotton, W.~D., Greisen, E.~W., et al.\ 1998, \aj, 115, 1693

\bibitem[Croom(2011)]{Croom2011} Croom, S.~M.\ 2011, \apj, 736, 161

\bibitem[Cutri et al.(2013)]{Cutri2013} Cutri, R.M.,  Wright, E.L., Conrow, T., et al. \ 2013,  \url{wise2.ipac.caltech.edu/docs/release/allsky/expsup/}

\bibitem[Decarli et al.(2011)]{Decarli2011} Decarli, R., Dotti, M., \& Treves, A.\ 2011, \mnras, 413, 39

\bibitem[Donato et al.(2001)]{Donato2001} Donato, D., Ghisellini, G., Tagliaferri, G., \& Fossati, G.\ 2001, \aap, 375, 739

\bibitem[Ferrarese \& Ford(2005)]{FerrareseFord2005} Ferrarese, L., \& Ford, H.\ 2005, \ssr, 116, 523

\bibitem[Frank et al.(2002)]{Frank2002} Frank, J., King, A., \& Raine, D.~J.\ 2002, Accretion Power in Astrophysics, by Juhan Frank and Andrew King and Derek Raine, pp.~398.~ISBN 0521620538.~Cambridge, UK: Cambridge University Press, February 2002

\bibitem[Ghisellini et al.(2010)]{Ghisellini2010} Ghisellini, G., Della Ceca, R., Volonteri, M., et al.\ 2010, \mnras, 405, 387

\bibitem[Giommi et al.(2012)]{Giommi2012} Giommi, P., Polenta, G., L{\"a}hteenm{\"a}ki, A., et al.\ 2012, \aap, 541, A160

\bibitem[Gold et al.(2011)]{Gold2011} Gold, B., Odegard, N., Weiland, J.~L., et al.\ 2011, \apjs, 192, 15

\bibitem[Graham et al.(2011)]{Graham2011} Graham, A.~W., Onken, C.~A., Athanassoula, E., \& Combes, F.\ 2011, \mnras, 412, 2211

\bibitem[Greene \& Ho(2005)]{GreeneHo2005} Greene, J.~E., \& Ho, L.~C.\ 2005, \apj, 630, 122

\bibitem[Haardt \& Madau(2012)]{HaardtMadau2012} Haardt, F., \& Madau, P.\ 2012, \apj, 746, 125

\bibitem[Hancock et al.(2011)]{Hancock2011} Hancock, P.~J., Roberts, P., Kesteven, M.~J., et al.\ 2011, Experimental Astronomy, 32, 147


\bibitem[Hinshaw et al.(2009)]{Hinshaw2009} Hinshaw, G., Weiland, J.~L., Hill, R.~S., et al.\ 2009, \apjs, 180, 225

\bibitem[Ivison et al.(2007)]{Ivison2007} Ivison, R. J., Greve, T.~R., Dunlop, J.~S., et al.\ 2007, \mnras, 380, 199I


\bibitem[Kaspi et al.(2000)]{Kaspi2000} Kaspi, S., Smith, P.~S., Netzer, H., et al.\ 2000, \apj, 533, 631


\bibitem[Kaspi et al.(2005)]{Kaspi2005} Kaspi, S., Maoz, D., Netzer, H., et al.\ 2005, \apj, 629, 61



\bibitem[Koratkar \& Gaskell(1991)]{KoratkarGaskell1991} Koratkar, A.~P., \& Gaskell, C.~M.\ 1991, \apjl, 370, L61

\bibitem[Laor (1990)]{Laor1990} Laor, A.\ 1990, \mnras, 246, 369

\bibitem[Li et al.(2005)]{Li2005} Li, L.-X., Zimmerman, E.~R., Narayan, R., \& McClintock, J.~E.\ 2005, \apjs, 157, 335

\bibitem[Malkan (1983)]{Malkan1983} Malkan, M. A.\ 1983, \apj, 268, 582

\bibitem[Mannucci et al.(2001)]{Mannucci2001} Mannucci, F., Basile, F., Poggianti, B.~M., et al.\ 2001, \mnras, 326, 745



\bibitem[Massaro et al.(2011)]{Massaro2011BZCAT} Massaro, E., Giommi, P., Leto, C., et al.\ 2011, Multifrequency Catalogue of Blazars (3rd Edition), Edited by E.~Massaro, P.~Giommi, C.~Leto, P.~Marchegiani, A.~Maselli, M.~Perri and S.~Piranomonte.~ARACNE Editrice, Rome, Italy, 118 pages



\bibitem[Morrissey et al.(2007)]{Morrissey2007} Morrissey, P., Conrow, T., Barlow, T.~A., et al.\ 2007, \apjs, 173, 682

\bibitem[Netzer(1990)]{Netzer1990} Netzer, H., 1990, agn..conf...57N

\bibitem[O'Donnell(1994)]{ODonnell1994} O'Donnell, J.~E.\ 1994, \apj, 422, 158

\bibitem[Onken et al.(2004)]{Onken2004} Onken, C.~A., Ferrarese, L., Merritt, D., et al.\ 2004, \apj, 615, 645

\bibitem[Park et al.(2012a)]{Park2012a} Park, D., Kelly, B.~C., Woo, J.-H., \& Treu, T.\ 2012a, \apjs, 203, 6

\bibitem[Park et al.(2012b)]{Park2012b} Park, D., Woo, J.-H., Treu, T., et al.\ 2012b, \apj, 747, 30


\bibitem[Planck Collaboration VII(2011)]{PlanckCollaborationVII2011} Planck Collaboration VII\ 2011, \aap, 536, A7


\bibitem[Planck Collaboration XIII(2011)]{PlanckCollaborationXIII} Planck Collaboration XIII 2011, A\&A, 536, A13

\bibitem[Polletta et al.(2007)]{Polletta2007} Polletta, M., Tajer, M., Maraschi, L., et al.\ 2007, \apj, 663, 81

\bibitem[Schneider et al.(2010)]{Schneider2010} Schneider, D.~P., Richards, G.~T., Hall, P.~B., et al.\ 2010, \aj, 139, 2360


\bibitem[Shakura \& Sunyaev(1973)]{ShakuraSunyaev1973} Shakura, N.~I., \& Sunyaev, R.~A.\ 1973, \aap, 24, 337

\bibitem[Shang et al.(2007)]{Shang2007} Shang, Z., Wills, B.~J., Wills, D., \& Brotherton, M.~S.\ 2007, \aj, 134, 294

\bibitem[Shankar et al.(2009)]{Shankar2009} Shankar, F., Weinberg, D.~H., \& Miralda-Escud{\'e}, J.\ 2009, \apj, 690, 20

\bibitem[Shaw et al.(2012)]{Shaw2012} Shaw, M.~S., Romani, R.~W., Cotter, G., et al.\ 2012, \apj, 748, 49

\bibitem[Shen et al.(2011)]{Shen2011} Shen, Y., Richards, G.~T., Strauss, M.~A., et al.\ 2011, \apjs, 194, 45

\bibitem[Shen(2013)]{Shen2013} Shen, Y., 2013, arXiv:1302.2643

\bibitem[Trushkin(2003)]{Trushkin2003} Trushkin, S.~A.\ 2003, Bulletin of the Special Astrophysics Observatory, 55, 90


\bibitem[Vestergaard \& Peterson(2006)]{VestergaardPeterson2006} Vestergaard, M., \& Peterson, B.~M.\ 2006, \apj, 641, 689

\bibitem[Wandel \& Petrosian (1988)]{Wandel_Petrosian1988} Wandel, A., \& Petrosian, V.\ 1988, \apj, 329, 11

\bibitem[Wright et al.(2010)]{Wright2010} Wright, E.~L., Eisenhardt, P.~R.~M., Mainzer, A.~K., et al.\ 2010, \aj, 140, 1868

\bibitem[Wu et al.(2004)]{Wu2004} Wu, X.-B., Wang, R., Kong, M.~Z., Liu, F.~K., \& Han, J.~L.\ 2004, \aap, 424, 793



\end{thebibliography}
\end{document}